# Coding limits on the number of transcription factors

Shalev Itzkovitz[1,2], Tsvi Tlusty[2], Uri Alon[1,2§]

[1]Depts. Molecular Cell Biology and [2]Physics of Complex Systems Weizmann Institute of Science, Rehovot 76100, Israel

Corresponding author

[§]Email addresses:

- SI: shalev.itzkovitz@weizmann.ac.il
- TT: tsvi.tlusty@weizmann.ac.il
- UA: uri.alon@weizmann.ac.il

# Abstract

**Background**

Transcription factor proteins bind specific DNA sequences to control the expression of genes. They contain DNA binding domains which belong to several super-families, each with a specific mechanism of DNA binding. The total number of transcription factors encoded in a genome increases with the number of genes in the genome. Here, we examined the number of transcription factors from each super-family in diverse organisms.

**Results**

We find that the number of transcription factors from most super-families appears to be bounded. For example, the number of winged helix factors does not generally exceed 300, even in very large genomes. The magnitude of the maximal number of transcription factors from each super-family seems to correlate with the number of DNA bases effectively recognized by the binding mechanism of that super-family. Coding theory predicts that such upper bounds on the number of transcription factors should exist, in order to minimize cross-binding errors between transcription factors. This theory further predicts that factors with similar binding sequences should tend to have similar biological effect, so that errors based on mis-recognition are minimal. We present evidence that transcription factors with similar binding sequences tend to regulate genes with similar biological functions, supporting this prediction.

**Conclusions**

The present study suggests limits on the transcription factor repertoire of cells, and suggests coding constraints that might apply more generally to the mapping between binding sites and biological function.

# Background

Transcription factor proteins regulate genes by binding DNA sequences at the promoters of the target genes. Typically, each transcription factor (TF) is able to recognize a set of similar sequences, centred around a consensus sequence [1-11]. The binding probability is generally believed to be higher the more similar a sequence is to the consensus sequence.

Transcription factor proteins can be classified into different super-families, each one with a different DNA-binding mechanism [12-16]. For example, the winged helix super-family consists of proteins which insert an alpha helix into the major groove of the DNA, forming amino-acid-base contacts over a region spanning about 5-6 base-pairs. These proteins tend to form homo-dimers, which often contact two consecutive major grooves [13]. Thus, their binding sequences are palindromic repeats of a 5-6 base-pair sequence. Proteins from the homeodomain-like super-family insert an alpha helix parallel to the DNA backbone, and tend to form heterodimers, thus forming more independent base-pair contacts than the winged helix proteins. Other super-families like the C2H2 zinc-coordinating super-family are monomer proteins with variable number of DNA recognition domains, or 'fingers', each recognizing 3 consecutive base-pairs [17, 18].

The total number of transcription factors (TFs) encoded by a genome increases with the number of genes in the genome. The number of TFs has been shown to scale with genome size as a power-law (the number of TFs, N, scales as the number of genes G as $N \sim G^{1.9}$ for Prokaryotes and $N \sim G^{1.3}$ for Eukaryotes [19]). This is thought to reflect the fact that the more complex the organism, the more intricate the regulation needed to respond to environmental inputs and to carry out developmental programs.

Here we ask whether there are limits on the numbers of transcription factors from different super-families. We find that the maximal numbers of transcription factors from most super-families is significantly smaller than the total number of transcription factors. The maximal number for each super-family appears to correlate with the number of possible sequences effectively recognized by the binding mechanism of that super-family. We also show that the binding sequences of different TFs are often overlapping, and that TFs with similar binding sequences tend to participate in similar biological processes. The results are compared to simple coding models that may provide an intuitive understanding of the origin and magnitude of the bounds on TF numbers.

## Results

**Maximal number of transcription factors in most super-families is bounded**

The total number of TFs increases with the number of genes in the genome [19], exceeding 2700 TFs in organisms such as *Xenopus tropicalis*. However, when considering each TF super-family separately, we find that the number of TFs from most super-families reach a maximum size which is significantly lower than the total number of TFs (in other words, the size of the super-family is bounded). For example, winged helix transcription factors increase with number of genes until reaching a maximum of about 300 TFs in organisms with ~5000 genes (Table I). Larger genomes contain winged helix TFs but at numbers which do not appear to exceed this bound.

A similar picture is observed in other super-families (Table I), for example, the maximal number of lambda-repressors is about 80 (Table I) and the maximal number of helix-loop-helix proteins is about 185 (Table I). One super-family - the multi-domain C2H2 zinc finger super-family, display a significantly higher maximal number of TFs. These proteins, found mainly in eukaryotes, increase in number with genome size, following genome size as about the number of genes squared ( $G^a$ with a=1.8+/- 0.17).

**The maximal number of TFs correlates with number of degrees of freedom in the binding mechanism**

In this section we compare the magnitude of the maximal TF numbers with the number of degrees of freedom in the binding mechanism of each super-family. The results are summarized in Fig 1 and Table I.

The number of degrees of freedom of a binding mechanism is related to the number of different base-pairs that can be specifically recognized by the DNA-binding mechanism. For example, lambda-repressor like proteins recognize DNA by inserting a short alpha helix into the major groove, specifically recognizing only three base-pairs [20, 21]. These three base-pairs essentially determine the binding sequence of the TF, because these proteins usually form homo-dimers in which each monomer recognizes essentially the same sequence [13]. The proteins from this super-family have a relatively constrained binding mechanism, with 64 possible binding sequences (since there are $4^3/2$ combinations of three bps, including reverse complement sequences, and there are two possible orientations of the half sites, see methods). The maximal number observed for proteins in this super-family is about 80 per genome (Table I).

Winged helix (wH) transcription factors recognize DNA using a similar mechanism of inserting an alpha helix into the major groove. However, the longer alpha helix used by these TFs usually interacts with 6 bp positions. Just as in lambda-repressor like TFs, these 6 base-pairs determine the binding sequence, because wH proteins usually form homo-dimers [13]. There does not appear to be any constraint on the possible 6 base pair sequence that can be recognized by a suitably designed wH protein. Hence, the maximal number of different sequences that can be recognized by such factors can be estimated as $4^6/2=2048$, more than the number of sequences available for lambda-repressor like TFs. The observed maximal number for this super-family, about 300 (Table I, Fig 1) is higher than the maximal number for the lambda-repressor like super-family.

Three other super-families have related mechanisms: helix-loop-helix proteins, Zn2/Cys6 proteins and glucocorticoid receptor-like proteins (Table I). These three super-families bind as dimers, in which each monomer binds a highly constrained sequence (half-site). Helix-loop-helix proteins usually recognize one of only a few conserved major-groove hexamer sequences, such as the E-box or G-box [22-24]. In these sequences, only two positions are variable. These proteins can form homo-dimers or hetero-dimers. This is the most constrained of the three super-families, and has the lowest observed maximal number, 185. Zn2/Cys6 proteins bind to three bp identical half-sites. They have more possible binding sequences than helix-loop-helix proteins, because the half-sites can be at variable spacing and orientations (estimated at ~1250 possible sequences vs ~130). The maximal number for this super-family is higher, 250 (table I, Fig 1). Glucocorticoid receptor-like proteins bind two half sites which can be at variable orientations and spacing [20] and in addition can form hetero-dimers that bind to non-identical half-sites. This super-family therefore has the most degrees of freedom of the three super-families (~3450 possible sequences), and displays the highest maximal number, about 380.

C2H2 proteins have between two to more than 30 finger domains, each recognizing three consecutive base-pairs[18, 25]. These proteins have the largest number of possible binding sequences ($64^n/2$ for an n-domain protein). The maximal number of such proteins in a single organism is the highest of all super-families, consistent with the large number of degrees of freedom for the possible binding sites.

Table I and Fig 1 show the maximal numbers for all super-families, as well as an estimate for the number of possible sequences where data on the mechanism is available. The maximal number seems to increase with the number of possible binding sequences.

**Evolutionary shift to super-families with more degrees of freedom**

When examining the distribution of transcription factors among the different super-families for different organisms (Fig 2) one can observe a shift from the predominant use of super-families with less degrees of freedom and smaller maximal numbers to super-families with more degrees of freedom and higher maximal number of TFs. Organisms with a small number of genes predominantly use TF super-families such as Lambda-repressor like and C-terminal effectors, while as organisms with more genes shift to Znz-Cys5 and Glucocorticoid receptor-like super-families, which have higher maximal numbers of TFs. Organisms with more genes, such as mouse and human, shift to the predominant use of C2H2 multi-domain zinc finger TFs, which have the highest maximal number.

**Coding theory suggests upper bounds for transcription factor numbers**

What is the origin of the bounds on the numbers of transcription factors? As one possible explanation, we consider the mapping of transcription factors to binding sequences as a coding problem, analogous to the assignment of amino acids to codons in the genetic code. We would like to emphasize that the purpose of models in this study is not to serve as descriptions of precise biochemical mechanisms, but rather as simple conceptual guides to understand the essential forces at play. Thus, the models neglect such details as whether each residue in the binding domain of the TF recognizes one or more DNA bases, as well as issues of DNA malleability, non-specific amino-acid base contacts, etc. The models also neglect important effects such as cooperative binding between TFs and other regulatory features.

To begin, consider a hypothetical situation in which each sequence would be assigned a different TF. In the case of winged helix TFs, for example, in which a binding sequence is effectively of length 6, there are in principle $4^6/2=2048$ different sequences (or 2080 if one considers separately the 64 self-complementary sequences in addition to $(4^6-64)/2$ unique non-self complementary sequences). There are therefore a maximal number of 2048 TFs that are perfectly stringent and recognize only one sequence. In reality, however, each TF recognizes a set of sequences, located around a consensus sequence [1-11]. Thus, the assignment of TFs to sequences should assign to each TF a set of adjacent sequences. This raises a difficulty, because TFs with very similar sets of binding sequences can recognize each others binding sequences.

In the theoretical case of perfectly non-overlapping sequences, in which each sequence is assigned to only one TF, Sengupta et al [6] have noted that the coding problem is similar to the problem of packing non-overlapping spheres in the space of sequences, each sphere corresponding to the sequences belonging to one TF (Fig 3a). Let us make a simple estimate of the number of TFs according to this picture. As an example, suppose that a TF can on average recognize sequences that are one letter different from the consensus (Hamming distance of one away from the consensus sequence). For winged helix proteins, there are six positions in the sequence, each of which can be changed to one of 3 other letters, resulting in 6x3=18 neighbours that are a Hamming distance of one away. Thus, there can be at most 2048/19~100 distinct TFs with non-overlapping sequences [26]. This is on the same order of magnitude, although lower than the observed maximal number of about 300 (Table I).

Coding theory suggests that one can increase the number of TFs by allowing sequences to overlap. This comes at a cost: TFs can mis-recognize each other's sequences leading to errors in gene expression. Optimal codes that can minimize such errors are known as "Gray codes" in information theory [27]. An optimal coding theory, which allows sequences to overlap, has been recently suggested in the context of the genetic code [28]. In the genetic code, codons differing by one base-pair correspond either to the same amino acid or to chemically related amino acids [29-31]. This mapping is thought to minimize the error load caused by errors in translation [28, 31].

Here we apply this theory to TFs. Since the theory takes into account the mis-binding errors, it can reach higher bounds than hard-sphere packing codes (Table II). Importantly, the theory predicts that neighbouring "spheres", that is TFs with similar binding sequences, would tend to be close in function in order to minimize the error load. Thus, the TFs with overlapping sequences should regulate the same genes, or genes with similar functions, so that effects of cross-recognition are minimized. Such codes are called "smooth" (Fig 3b).

**Factors with similar binding sequences tend to have similar biological functions**

A qualitative prediction of the smooth coding theory is that factors with similar recognition sequences should tend to have similar biological effects. The reason is that factors with similar sequences can sometimes bind to each others sequences. If the biological effects of binding of TF A and B are similar, the reduction in fitness caused by such errors would be smaller. Hence, there may be a selection pressure to allocate similar sequences to biologically similar factors.

To test the prediction that TFs with similar sequences should tend to have similar function, we examined TFs in *E. coli*, yeast and human, and compared their sequence similarity by means of several distance metrics. In these organisms there exists a significant sequence similarity between the binding sites of some TF pairs (Fig 4-6). The yeast set of 94 well characterized TFs contained 18 pairs with highly similar sequences (Fig 4). The *E. coli* set of 46 well characterized TFs contained 6 pairs with highly similar sequences (Fig 5). The human set of 49 TFs contained 9 pairs with highly similar sequences (Fig 6). In other words, the TF "spheres" often overlap significantly (Fig 3B).

To assess the similarity in function of the TFs with similar binding sequences, we used two measures. The first similarity measure was a significant co-regulation of target genes by both factors. The second measure was the similarity in functional annotation [32, 33] of the target genes of each TF. For both measures, we observed a significant enrichment of TF pairs with similar sequences and similar biological function measures.

We now provide more details on this result. To assess the functional similarity in yeast, we used an experimentally determined transcription network [34, 35]. This network contained targets for 64 TFs in our data-set. About 14% (276/2016) of all TF pairs had significant target co-regulation. When considering pairs with similar binding sequences, the fraction with significant target co-regulation increases to over 50% (10/18, p-value of $5.1*10^{-5}$) (Fig 4).

As a second measure of functional similarity, we assigned to each yeast TF a profile according to the functional annotation of its target genes [32]. We then compared the average distances between the profiles of TF pairs with similar binding sequences to the average distances between the profiles of all TF pairs. We find that TF pairs with similar binding sequences have a lower average profile distance (0.17 vs 0.35, p-value of $7*10^{-5}$). Thus their targets tend to have more similar biological annotations. Examples of such pairs are the stress-response regulators MSN2 and MSN4, the stress-response YAP TFs, the cell cycle regulators FKH1 and FKH2 and the nitrogen regulators GLN3 and DAL82.

Similar results were obtained for *E. coli*. In our database of 46 TFs we found 6 pairs of TFs with significantly high binding sequence similarity (Fig 5). To assess the functional similarity between these pairs, using co-regulation criterion, we parsed the Ecocyc database [33] to obtain a network of 541 operons and 806 experimentally verified transcription interactions. Only 4% (39/1035) of all TF pairs had significant target co-regulation. When considering TFs with similar binding sequences, this fraction increases to over 65% (4/6, p-value $2.5*10^{-5}$). As a second test, we used a functional annotation for *E. coli*. We found that TF pairs with similar sequences have a lower average functional profile distance (0.26 vs 0.76, p-value of $4.2*10^{-5}$). Thus, the target genes of these TFs tend to have similar biological annotations. Examples of TF pairs with similar sites and similar functions include the drug and stress response regulators MarA, SoxS and Rob that jointly regulate at least 6 operons, and the anaerobic metabolism regulators NarL and NarP that jointly regulate 5 operons. The similarity in sequences is so large that some of these factors bind the exact same sequence in some of their co-regulated genes. However, many of their binding sequences are also distinct (the TF spheres appear not to overlap completely).

For human, limited data currently prohibits a detailed statistical analysis. However, several examples are known where functionally related TFs recognize the same sequences. These include the interferon regulatory factors IRF-1 and IRF-2 [36], and the ETS transcription factors SPI-B and SPI-1 [37].

Many TF pairs with overlapping sequences and similar function are close paralogs (belong to the same family within the super-family) [15, 38-40]. In most cases there are additional paralogous TFs with non-similar binding sequences. For example the *E. coli* regulators MarA, SoxS and Rob that recognize similar sequences, are all paralogs from the AraC/XylS family. However, they differ in their binding sequences and in their biological function from their paralog AraC. This suggests that homology of TF proteins does not fully explain the similarity in their binding sites. Gene duplication may aid in generating paralogous TFs with similar binding sites, which can then be selected according to the cost and benefit of their action on the target genes.

## Discussion

The main result of this study is that the maximal numbers of TFs from most transcription factor super-families appear to be bounded. The number of these TFs in a genome does not seem to exceed a certain upper bound. These bounds range from around 80 for lambda repressor-like, to about 420 for homeodomain proteins. The bounds seem to correlate with the number of degrees of freedom of the DNA-binding mechanism in each super-family. The multi-domain C2H2 zinc fingers super-family displays a significantly higher maximal number in the present data, compared to other super-families.

To understand these bounds, we considered the coding problem faced by the cell: how to assign different sequences to each transcription factor in a way that avoids erroneous recognition in which a transcription factor binds where it shouldn't. As organisms of increasing complexity evolve there is a need for more diversity in gene regulation, through the introduction of new transcription factors. The stochastic process of DNA recognition by TFs may result in binding of a TF to binding sequences intended for another TF, if these are similar enough. This study examined the proposal that minimizing these misrecognitions limits the maximal numbers of TFs with a given binding mechanism, that is, TFs from a given super-family.

To examine the coding problem on a qualitative level, we considered two theoretical mappings of code-words (DNA sequences) to messages (TFs): a sphere packing code in which each TF has unique sequences not shared with other TFs, and a smooth code in which sequences of different TFs can partially overlap. The latter appears to offer a more realistic representation, because many pairs of TFs have highly overlapping code-words. For some super-families, the observed bounds seem to approximately agree with the theoretical bound derived for smooth codes.

A prediction of the theory in which misrecognition errors are an important constraint on the coding, is that TFs with similar binding sequences should tend to have similar biological functions. This is because misrecognition between such TFs would have a smaller impact on fitness. This prediction agrees with the present observation that the TFs with overlapping code-words are significantly closer in their biological function than expected at random.

One possible scenario for the evolution of TF super-families is as follows. Simple organisms, which require few TFs, employ certain super-families such as lambda repressor-like and winged helix. When these super-families reach their maximal limits, new super-families are needed. At these points organisms shift their TF usage to novel super-families with more degrees of freedom and higher maximal numbers (Fig 2). An example is the increased use of the C2H2 zinc finger TFs in the more advanced organisms.

It is important to note that the usage of different TF super-families is linked to the phylogenetic grouping of organisms. An example is the Zn2/Cys6 DNA-binding domain TFs which are largely restricted to fungal organisms [14, 41]. While as the increase in the required number of TFs coupled with the coding limits suggested here may create an evolutionary pressure for the evolution of new TF families, this phylogenetic grouping can still be observed in the TF distributions (Fig 2).

Much of the innovation of biological function has been attributed to events of gene duplications [42-45]. Several species, such as zebra-fish and Arabidopsis have passed whole genome duplications. Such duplication events may lead to a situation where the number of TFs from a given super-family exceeds its theoretical bound. This could either create an evolutionary pressure which may lead to rapid loss of these genes or result in a redundant, non-equilibrium census of TFs. As an example of the first scenario, zebra-fish is thought to have passed a whole genome duplication in the lineage leading from its last ancestor with humans [44]. The number of helix-loop-helix TFs in zebra-fish is approximately the same as in humans (149 compared to 164).

To further test the present conclusions requires additional biological data. The current sets of TFs with known binding sequences (the transcriptional code) and gene-regulation networks, representing the functional annotation of the TFs, are still partial. For example, the current dataset for *E. coli* includes less than a fifth of all TFs in the organism. Once datasets are enlarged, one may get a better estimate of the bounds, the amounts of overlap in sequence space, and the functional smoothness of the transcriptional code.

Along the same lines, further knowledge of TF-DNA binding mechanisms could allow one to obtain more accurate estimates for the number of degrees of freedom of each super-family in order to more accurately test the correlation of the observed bounds with this number. In the case of C2H2 zinc fingers, detailed structures of the TF-DNA complex allowed a good estimation of the mapping between residues in the TF binding domain and the DNA bases recognized by the TF [17, 18]. In other super-families, however, no clear mapping has yet been devised. Therefore, the present study presented only crude estimates for the number of possible sequences of each super-family. The present study predicts that the number of degrees of freedom in super-families with a lower observed bound should be smaller than for super-families with high bounds.

The present study focused only on one level of TF-DNA interaction, the recognition mechanisms of DNA binding sequences by transcription factors. There are many additional effects that govern transcription regulation. Gene expression often depends on the combined effects of multiple TFs, integrated by a cis-regulatory input function at the promoter [46-53]. The functional role of each TF is influenced by the distance on the promoter of its binding sites from sites of other TFs and the transcription start site [49], as well as the phase of the site along the DNA helix [54] [55]. In addition, there is often co-operative binding to other factors [56, 57], tissue-specific TF expression [58] and differential exclusion of TFs from the nucleus that is dependent on cell type and conditions [59]. These have been proposed as explanations for the smaller power-law exponent observed in the scaling of the total number of TFs with the number of genes in eukaryotes relative to prokaryotes [19]. By effectively introducing more degrees of freedom into the binding mechanism, these additional effects may also alleviate the constraints anticipated by the high levels of sequence overlap observed in the present study. These effects may explain the abundance of TFs from super-families like the lambda repressor-like and helix-loop-helix TFs, for which the observed maximal number of TFs are higher than the expected bounds in the simplest coding models.

The creation of maximal diversity of TFs with minimal misrecognition error-load might not be the only factor underlying the smooth codes observed in this study. Assigning TF pairs with similar biological function to similar binding sequences may have additional functional advantages. Some of the TF pairs with overlapping binding sequences and similar biological function presented in this study, such as the yeast MSN2 and MSN4 stress response regulators, are partially redundant. Other TF pairs, such as the IRF-1 and IRF-2 in humans have an antagonistic regulation mode, where one activates and another represses the same target genes on different time-scales [36]. This kind of regulation may create a transient activation profile, where target genes are activated for a short time following induction. TF redundancy [60] and antagonistic regulation may form additional 'forces' pulling TF sequence sets together, increasing the number of TFs above the strict coding bounds.

## Conclusions

In conclusion, the present study suggests that there are upper bounds on the number of transcription factors from different super-families. It seems that the more constrained the binding mechanism, the lower the bound. The present bounds may be understood in terms of an optimal coding strategy, in which misrecognition errors are minimized. As predicted by such a theory, the present data suggests that TFs with similar binding sequences tend to regulate genes with similar biological functions. More generally, similar coding problems may occur in other recognition problems in biology, such as protein-RNA recognition and protein-protein interactions through defined protein recognition motifs[61, 62]. Coding constraints can potentially limit the number of different protein binding motifs of a given type in the cell, in order to avoid non-specific cross recognition. It would be interesting to extend the present approach to these and other molecular recognition systems.

## Methods

**Transcription factor numbers**

We focused on ten major super-families of transcription factors: lambda repressor-like, C-terminal effector domain of the bipartite response regulators, winged helix, srf-like domains, DNA binding domain (GCC box), helix-loop-helix, Zn2/Cys6, glucocorticoid receptor-like (hormone receptors), C2H2 and C2HC zinc fingers and homeodomains (Table I). We used the superfamily database (version 1.69) to obtain the numbers of TFs from each super-family in different organisms. The superfamily database contains extensive annotations of structural domains of proteins in 250 sequenced organisms using Hidden Markov Model profiles [63]. The database contains 1439 super-families. We focused on the major super-families of transcription factors studied in [15], and added all super-families that contained the terms "DNA binding" or "transcription". This resulted in 32 super-families. We further filtered super-families in which the maximal number of predicted proteins in a single organism was smaller than 50. For the remaining 10 super-families, we determined the maximum number of TFs as the maximal number of proteins from each super-family after discarding organisms with less than 5 proteins and discarding the top 1% of the remaining organisms. It is important to note that the super-family domain assignment may contain predicted transcription factors due to the appearance of the relevant structural domain, which are in fact not functional, or which have other roles in the cell [64]. Thus, the maximal numbers presently found may be an overestimate.

**Binding sequence databases**

Position-Specific Score Matrices (PSSM) for 46 *E. coli* transcription factors, were constructed based on the RegulonDB database [65]. The set of transcription factor binding sequences for each TF were searched for aligned motifs using AlignACE [66]. We chose the top-scoring motif, and considered only TFs with four or more aligned sequences contributing to that motif. Finally we removed the non-specific DNA binding factors FIS, HNS and IHF [20].

For the yeast *S. cerevisiae*, we used 94 PSSMs based on a set of 102 PSSMs constructed by Harbison et al [9]. We filtered out proteins that either do not bind DNA directly or always bind as a complex: Gal80, DIG1, STB1, Met4, HAP2, 3, 4, 5. All PSSMs were converted to a probability representation, where the sum of each PSSM column is 1.

For humans we used the PSSMs in the JASPAR database [67]. This data set consisted of 49 PSSMs.

**Measurement of sequence similarity**

To measure the similarity between binding sequences of a pair of factors we assessed the distances between their PSSMs. To compare pairs of PSSMs, we use a distance measure related to the one used by Wang et al [68]. The present measure, described below, is stringent in the sense that it scores bases according to their conservation within the PSSM, and compares to randomized PSSMs that preserve these conservation profiles. It is more appropriate for the present purpose than simpler methods such as direct comparison of sequence Hamming distances, because the present interest is in the active base pairs in the site, rather than base pair differences that have little functional impact on binding.

We denote the length of the binding sequence for $TF_i$ as $n_i$, and its PSSM by $M_i$. The PSSM is an $4*n_i$ matrix in which each column, $p_{i,k}$ is a vector of length 4 holding the probability of observing letters A,C,G,T at position $k$ in the set of aligned binding sequences of $TF_i$. For each PSSM we created an information profile [7] denoted by $I_i$. The information profile is a $n_i$ length vector whose k'th element quantifies the conservation of position $k$ in the PSSM:

(1) $I_{i,k} = 2 - H(p_{i,k})$

where

(2) $H(p_{i,k}) = -\sum_{m=1}^{4} p_{i,km} \log_2(p_{i,km})$

is the entropy of $p_{i,k}$. $I_{i,k}$ has a minimum of 0 when all four bases have equal probability of appearing at position $k$, and a maximum of 2 when all aligned binding sequences have the same base at position $k$ (the small sample size correction of [7] was applied). As the PSSMs in our database have different lengths, and the 'important' positions for the TF-target recognition are the conserved positions, we used the information profiles $I$ as weight vectors when comparing two PSSMs.

We define the similarity between $TF_i$ and $TF_j$ as:

(3) $D_{ij} = \max_{A}(\sum_{k \in A} I_{i,k} \cdot I_{j,k} \cdot d_{ij,k})$

The maximum is taken over all relative shifts $A$ of the two PSSMs, with a minimum of 5-positions of overlap. $d_{ij,k}$ is a similarity of the k'th position of the relatively shifted PSSMs. We used two different measures for $d_{ij}$ : One minus the Shannon-Jensen distance [69]:

(4) $d_{ij,k} = 1 - [H(\frac{p_{i,k} + p_{j,k}}{2}) - \frac{H(p_{i,k}) + H(p_{j,k})}{2}]$

and a simple correlation between the two probability vectors $p_{i,k}$ and $p_{j,k}$. Both measures gave very similar results. For each pair we computed the similarity with the reverse complement as well and took the maximal similarity. To detect pairs with significant similarity we first chose only pairs for:

(5) $D_{ij} > f * \min(D_{ii}, D_{jj})$

Where $f$ is a numerical factor (we used $f=0.75$). This amounts to requiring that the similarity between the sets of binding sequences of two TFs comparable to the similarity between the sequences of each TF. For the pairs that passed this criterion, we created an ensemble of 1000 random PSSM pairs and computed a distribution of similarities $D^r_{ij}$. The random PSSMs were created by randomly exchanging the A-T and C-G positions in each column of the original PSSMs. This operation preserves the information profile, as well as the GC content, and therefore forms a stringent ensemble. Similar sequences were sequences which had a p-value<0.005 for $D_{ij}$ using both distance measures.

**Measurement of similarity of biological function of TFs**

We defined two pairs of transcription factors as functionally similar if they jointly co-regulate a significant number of target genes. This information was obtained from transcription regulation networks: For yeast we used the network of [34, 35]. For E. coli we used the network based on the data in the Ecocyc database [33]. It is important to note that these networks are based on direct experimental measurements, and not on putative interactions based on binding site predictions. For each pair of TFs we used a hyper-geometric test to assess whether the number of genes regulated by both TFs is significantly larger than expected from the fraction of target genes of each TF alone. This measure of functional similarity normalizes for the variable number of target genes for different TFs. We used a hyper-geometric test to detect enrichment of TF pairs with significant target co-regulation in the group of TF pairs with similar sequences.

A second measure of biological similarity was based on the similarity of the functional annotation of the gene targets of each factor. We used functional annotations[32, 33] for the top tier of the annotation tree (except sub-cellular localization and general annotations such as "protein with binding function"). For yeast we used the following functional categories from the FunCat database [32]: metabolism, energy, cell cycle and DNA processing, cell rescue, defense and virulence, interaction with the cellular environment, interaction with the environment, transposable elements, viral and plasmid proteins, cell fate, development, biogenesis of cellular components, cell type differentiation.

For *E. coli* we used the following functional categories from the Ecocyc physiological roles annotation [33]: carbon utilization, degradation of macro-molecules, energy metabolism – carbon, energy production/transport, biosynthesis of building blocks, biosynthesis of macromolecules (cellular constituents), central intermediary metabolism, metabolism of other compounds, cell division, cell cycle physiology, motility, chemotaxis, energytaxis (aerotaxis, redoxtaxis etc), genetic exchange/recombination, adaptations, protection, defense/survival, DNA uptake. Each TF was assigned a profile vector in which each position holds the fraction of its target genes with the respective functional annotation. We used a student t-test to compare the average of the Euclidean distances between profile vectors of TF pairs with significant sequence similarity to the average of the Euclidean distances between the profile vectors of all pairs. Lack of data in humans currently prohibits a systematic measure of functional similarity of TFs.

**Assessment of the number of possible sequences**

We considered three features in the binding mechanism of each super-family which contribute to the number of possible sequences: the number of variable positions in each half-site, the relative spacing and orientation between them, and whether the sites are identical (for homo-dimeric TFs) or not (hetero-dimeric TFs). The number of possible sequences for a super-family with P variable positions in each half-site, O possible half-site orientations and S possible half-site spacings is $4^{P*H}*O*S/2$, where H=1(2) if the super-family binds as homo-dimers (hetero-dimers) respectively. In our calculations we divide by 2 to account for reverse complementary sequences. The present study presents these estimates for six of the ten TF super-families (Table I) for which data on these three features were available.

For the Lambda repressor-like super-family, previous work suggested 3 variable positions in each half site of the homo-dimer [20, 21] and 2 relative orientations of the two monomers [70], resulting in $4^3*2/2=64$ possible sequnces. Helix-loop-helix proteins can bind as either homo-dimers or hetro-dimers, and have two variable positions at each half-site, resulting in an estimated number of possible sequences of $4^4/2=128$. Zn2/Cys6 proteins, such as the yeast GAL4 protein, bind two sequences of length three, with variable spacing (0-12 nucleotides) and three possible orientations (direct, inverted or everted repeats) [41], resulting in $4^3*3*13/2=1248$ possible sequences. The glucocorticoid receptor-like proteins have two variable positions at each half site [20], and can bind as hetero-dimers in three different orientations (direct, inverted or everted repeats) and variable spacing ranging from 0-8 [71], resulting in $4^4*3*9/2=3456$ possible sequences. Winged helix and homeo-domain recognize 6 positions as either homo-dimers or hetro-dimers respectively, with a constant relative orientation and half-site spacing, resulting in 4096/2 and $4096^2/2$ possible sequences, respectively. An estimation of the number of possible sequences for multi-domain C2H2 zinc fingers is not given, as these proteins can contain a variable number of finger domains (between 2 and more than 30) per protein.

**Coding theory bounds - coloring number bound**

We treated the mapping between transcription factors and binding sequences as a coding problem, where the code-words are short DNA sequences of a given length and the messages are TFs. The code-words are abstracted as nodes of a graph, where edges connect any two code-words which differ by one base-pair (Hamming distance 1). This scenario has some similarities with the genetic code, where the code-words or "codons" are 3-base-pair strings, and the messages are the amino acids encoded by these codons. The error-load is the reduction in organism fitness due to erroneous binding of factor A to the code-word assigned to factor B. At one extreme, minimal error-load can be achieved by mapping all code-words to a single transcription factor. At the other extreme, maximal diversity is achieved by mapping each code-word to a different transcription factor, resulting in the same number of TFs as possible code-words.

If one assigns a biological function to the TFs that bind each code-word, it is clear that minimization of error-load would tend to smoothen this mapping. That is, TFs that are likely to bind similar sequences should have a similar biological function. For every mapping of binding sequences to factors, one can assign an error load, which measures the average impact of erroneous binding. It has recently been proposed [28] that in the limit of large errors, the maximal number of coded messages is bounded by the coloring number of the minimal surface which can embed the code-word graph [28]. Heawood's formula states that the coloring number is:

(6) $chr(\gamma) = int[\frac{1}{2}(7 + \sqrt{1 + 48\gamma})]$

where int[x] denotes the largest integer not greater than x. The coloring number depends on $\gamma$, the genus of the surface embedding the graph:

$$(7)\ \gamma = 1 - \frac{1}{2}(V - E + F) = 1 - \frac{1}{2}V(1 - \frac{d}{4})$$

Here $V = q^n/2$ is the number of possible code words encoded by a q-letter string of length n, assuming a code-word which is the reverse complement of another is not available for independent assignment. $E = V*(d/2)$ is the number of graph edges, and $F = V*(d/4)$ is an estimate of the number of faces of the surface embedding the graph. $d = (q-1)n$ is the number of neighbors of each code-word (by identifying sequences and their reverse complements each code-word has twice as many neighboring code-words as a code without the reverse-complement constraint but about half of these code-words are not available for independent assignment). Using this we get the following estimate for the coloring number:

$$(8)\ B(n,q) = chr(\gamma) \cong 3.5 + \sqrt{0.75q^n * [n(q-1) - 4]}$$

The bound for codes with different n are shown in Table III.

**Coding theory bounds - sphere packing bound**

An alternative possibility for the code mapping is one in which every sequence is assigned to only one transcription factor [6], and the probability of a misread error is thus negligible. The target DNA binding sequences of each transcription factor can be represented as a sphere in the code-word space (Fig 3). The center of each sphere is the consensus sequence, and all sequences differing from the consensus sequence by *e* positions (Hamming distance *e* from the consensus sequence) are assumed to be bound with a non-negligible probability by the TF[2, 8, 10]. Here we assumed *e=1*.

Unlike the smooth code, the spheres here are non-overlapping. The volume of a "sphere" of radius *e*, which contains all code-words with Hamming distance of *e* or less from a given code-word, is [72]:

$$(9)\ U(n,q,e) = \sum_{i=0}^{e} \binom{n}{i}(q-1)^i$$

The number *N* of non-overlapping spheres of radius 1 is bounded by:

$$(10)\ N \leq \frac{1}{2}A_q(n,1)$$

where:

$$(11)\ A_q(n,1) = \left\lfloor \frac{q^n}{U(n,q,1)} \right\rfloor$$

The upper bound in (11) is called the sphere-packing bound [26]. Codes which achieve this bound are called "perfect codes". In such codes the code-word space is fully covered by non-overlapping spheres of Hamming radius 1. Generally, the number of non-overlapping spheres is smaller, as some code-words remain uncovered by any sphere. The factor of one half in Eq. (10) stems from the fact that each sequence effectively represents also its reverse complementary sequence [26].

## Authors' contributions

All authors participated in the design of the study and writing of the manuscript. SI performed all the computational aspects of the work. All authors read and approved the final manuscript.

## Acknowledgements

We thank Nicholas Luscombe, Sarah Teichmann, Hannah Margalit, Naama Barkai, Eran Segal, Tzachi Pilpel, Michal Lapidot, Yael Garten and all members of our lab for useful discussions. SI acknowledges support from the Horowitz Complexity Science Foundation.

# Tables

**Table 1**

Maximal numbers of transcription factors from each super-family in a single organism, and the organism in which the maximum is observed. The kingdom in which each super-family is observed is abbreviated as A – Archea, B – Bacteria, E – Eukaryotes. Estimates for the number of possible sequences are shown (see methods). P – number of variable positions in each half-site, S – number of possible spacing between half-sites, O – number of possible orientations, H – homo-dimers (1) or hetero-dimers (2). The number of sequences is $4^{P*H}*O*S/2$.

|   | Super-family | Maximal # TFs | Kingdom | organism | P | S | O | H | # sequences |
|---|---|---|---|---|---|---|---|---|---|
| 1 | lambda repressor-like DNA-binding domains | 77 | A,B,E | *Photorhabdus luminescens* | 3 | 1 | 2 | 1 | 64 |
| 2 | C-terminal effector domain | 88 | A,B,E | *Streptomyces avermitilis* | - | - | - | - | - |
| 3 | srf-like | 122 | E | *Arabidopsis thaliana* | - | - | - | - | - |
| 4 | helix-loop-helix DNA-binding domain | 186 | E | *Arabidopsis thaliana* | 2 | 1 | 1 | 2 | 128 |
| 5 | DNA-binding domain | 194 | B,E | *Oryza sativa* | - | - | - | - | - |
| 6 | Zn2/Cys6 DNA-binding domain | 246 | E | *Fusarium graminearum* | 3 | 13 | 3 | 1 | 1,248 |
| 7 | winged helix DNA-binding domain | 299 | A,B,E | *Bordetella bronchiseptica* | 6 | 1 | 1 | 1 | 2,048 |
| 8 | glucocorticoid receptor-like DNA-binding domain | 376 | A,B,E | *C.elegans* | 2 | 9 | 3 | 2 | 3,456 |
| 9 | homeodomain-like | 417 | A,B,E | *Danio rerio* | 6 | 1 | 1 | 2 | $8.4*10^6$ |
| 10 | multi-domain C2H2 zinc fingers | 1308 | E | *Mus musculus* | 6-30 | 1 | 1 | 1 | - |

**Table 2**

Theoretical bounds for an n-length 4-letter code. The sphere packing bounds are from [25]. The coloring bound is given by equation (8).

| n | # code words - $4^n/2$ | Coloring bound | Sphere packing bound |
|---|---|---|---|
| 3 | 32 | 18 | 1 |
| 4 | 128 | 42 | 8 |
| 5 | 512 | 95 | 26 |
| 6 | 2,048 | 210 | 107 |
| 7 | 8,192 | 460 | 372 |
| 8 | 32,768 | 994 | 1310 |

# Figures

**Fig. 1.** Correlation between the maximal number of transcription factors and number of possible sequences for six super-families, for which details of binding mechanism are known.

**Fig. 2.** Distribution of transcription factors for the 10 organisms in Table 1 among the different super-families. On the x axis are the 10 super-families of table 1, on the y axis their counts in each organism. The organisms are sorted according to increasing number of genes in the genome.

**Fig. 3.** Conceptual coding schemes for the assignment of binding sequences to TFs. Binding sequences are displayed as points, TFs as colored spheres. Colors correspond to the biological function of each TF. a) A sphere-packing code – code-words are covered by non-overlapping spheres. The TFs do not share binding sequences. b) A smooth code - code-words are covered by overlapping spheres with similar function. TFs can share binding sequences with neighbor TFs. This type of code is predicted to be smooth, that is where TFs with shared binding sequences tend to have similar biological function, represented by spheres of similar color in the figure.

**Fig. 4.** Transcription factors with overlapping binding sequences in *S. cerevisae*. Nodes represent TFs, edges connect pairs of TFs if their corresponding sets of binding sequences have significant overlap according to the present measure. Bold edges connect TFs which also have biological similarity according to the functional annotation and transcription network (gene co-regulation) measures. Shown are the TF logos [11]. Logo length was limited to the highly conserved base pairs for clarity.

**Fig. 5.** Transcription factors with overlapping binding sequences in *E. coli*. Nodes represent TFs, edges connect pairs of TFs if their corresponding sets of binding sequences have significant overlap according to the present measure. Bold edges connect TFs which also have biological similarity according to the functional

annotation and transcription network (gene co-regulation) measure. Shown are the TF logos [11]. Logo length was limited to the highly conserved base pairs for clarity.

**Fig. 6.** Transcription factors with significantly overlapping sequences in Humans. Edges connect two TFs with similar binding sequences. Sequence logos are shown for each TF.

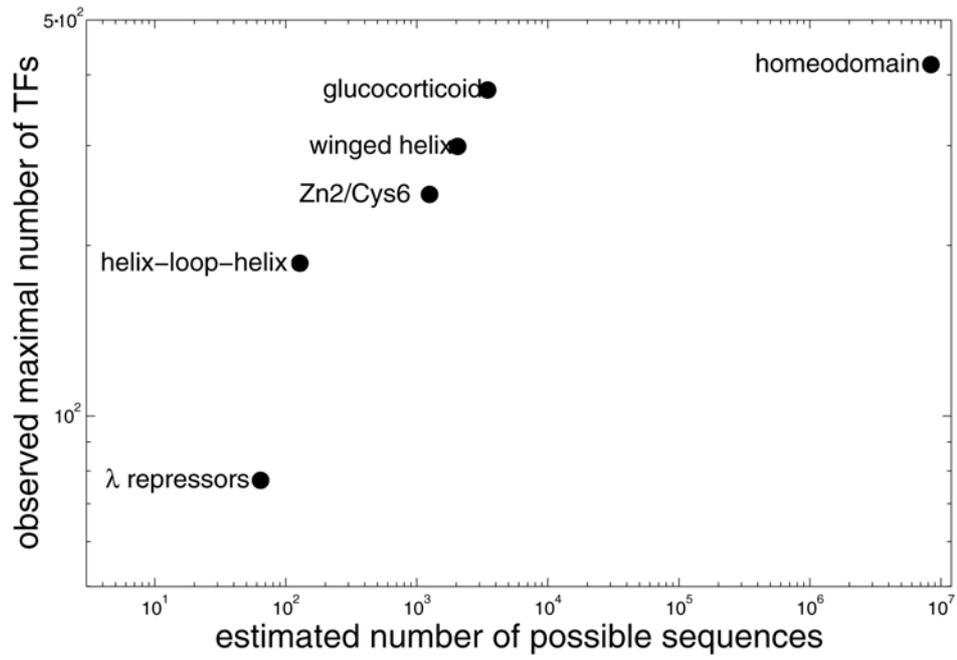

**Figure 1**

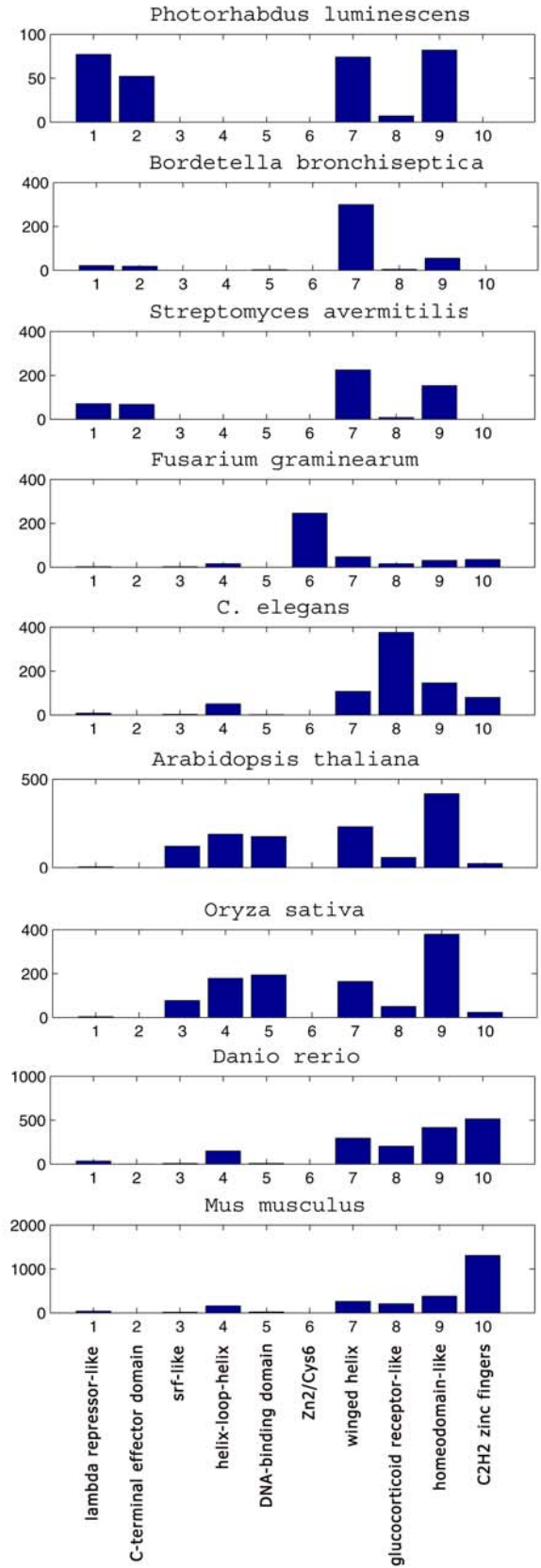

**Figure 2**

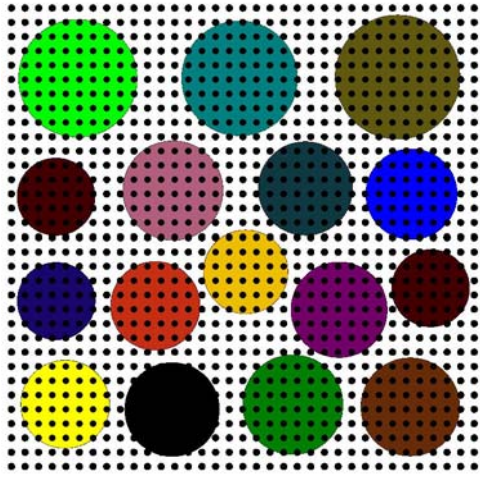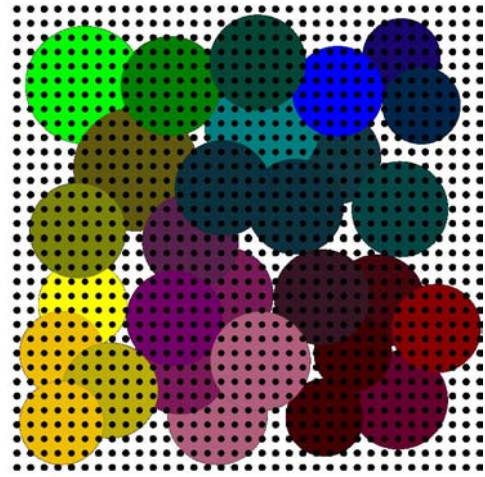

**Figure 3**

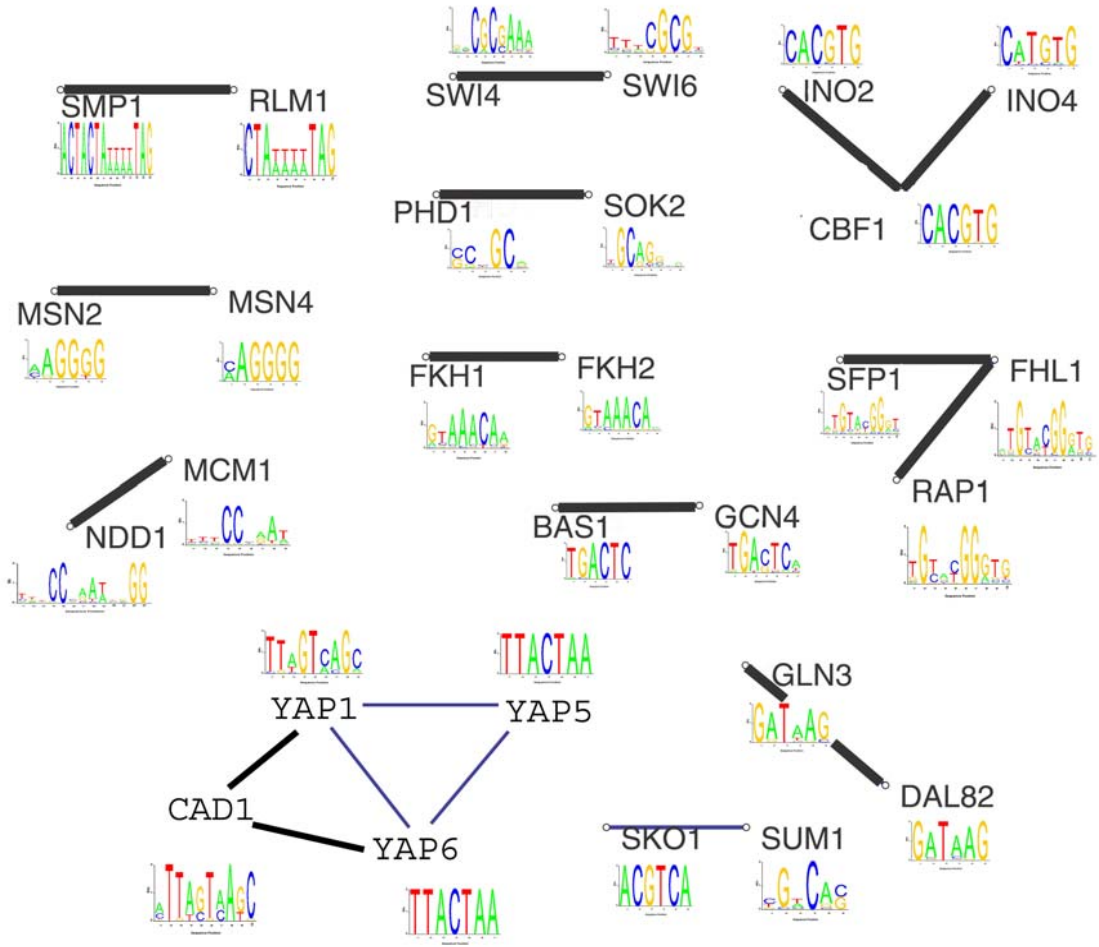

**Figure 4**

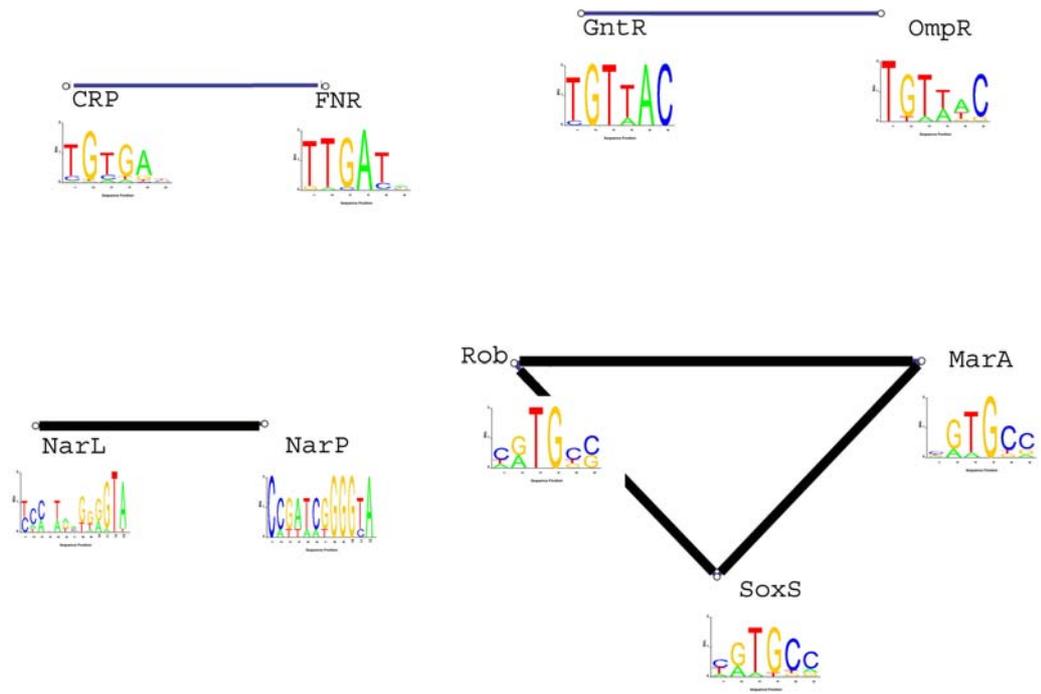

**Figure 5**

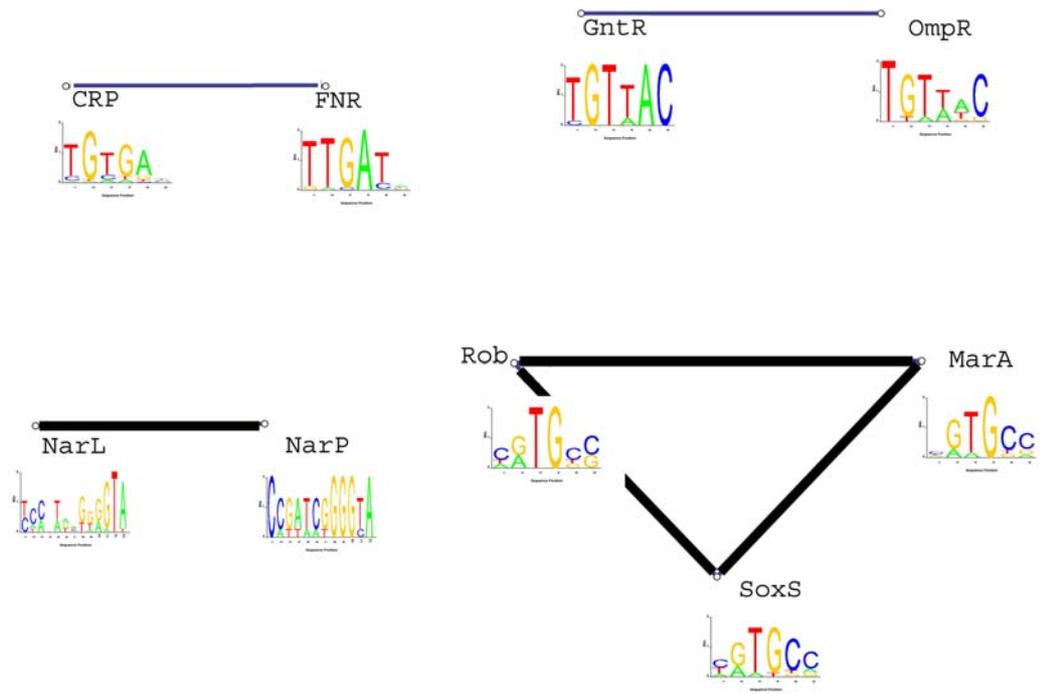

**Figure 5**

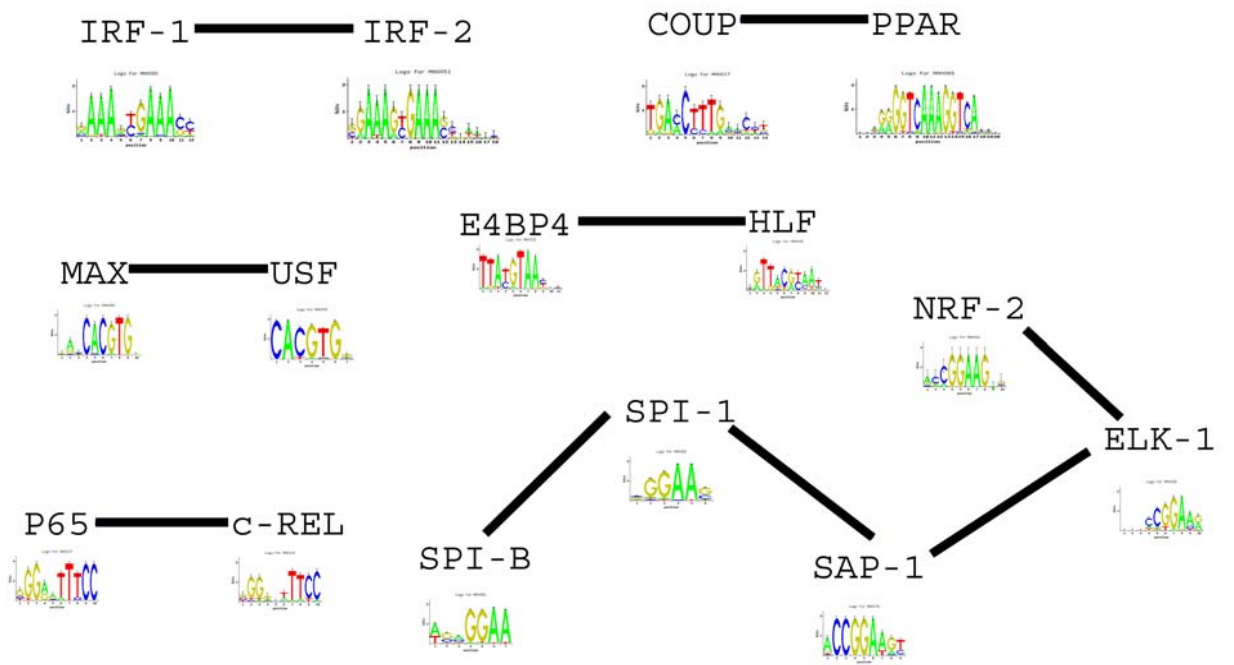

**Figure 6**